\documentclass[10pt,conference,letterpaper]{IEEEtran}
\usepackage{times,amsmath,epsfig}
\usepackage{multirow} 
\usepackage{flushend} 

\IEEEoverridecommandlockouts 
\IEEEpubid{\makebox[\columnwidth]{\copyright~2015~IEEE} \hspace{\columnsep}\makebox[\columnwidth]{ }}
\title{A horizontally-scalable multiprocessing platform based on Node.js}
\author{%
  {Ahmad Maatouki{\small $~^{\#1}$}, J\"org Meyer{\small $~^{\#2}$}, Marek Szuba{\small $~^{\#3}$}, Achim Streit{\small $~^{\#4}$} }%
\vspace{1.6mm}\\
\fontsize{10}{10}\selectfont\itshape
$^{\#}$\,Steinbuch Centre for Computing (SCC), Karlsruhe Institute of Technology (KIT),\\
Eggenstein-Leopoldshafen, Germany\\
\fontsize{9}{9}\selectfont\ttfamily\upshape
$^{1}$\,ahmad.maatouki@student.kit.edu\\
$^{2}$\,joerg.meyer2@kit.edu\\
$^{3}$\,marek.szuba@kit.edu\\
$^{4}$\,achim.streit@kit.edu%
\vspace{1.2mm}\\
\fontsize{10}{10}\selectfont\rmfamily\itshape
\fontsize{9}{9}\selectfont\ttfamily\upshape
}
\begin{document}
\maketitle
\begin{abstract}
  This paper presents a scalable web-based platform called Node Scala which
  allows to split and handle requests on a parallel distributed system
  according to pre-defined use cases. We applied this platform to a
  client application that visualizes climate data stored in a NoSQL
  database MongoDB. The design of Node Scala leads to efficient usage
  of available computing resources in addition to allowing the system
  to scale simply by adding new workers. Performance evaluation of
  Node Scala demonstrated a gain of up to 74~\% compared to the
  state-of-the-art techniques.

\end{abstract}

%

\section{Introduction}
Web applications offer the advantage to end users to be able to run applications inside browsers without the need to install software locally. If the web application retrieves input data from web sources users do not even need to explicitly manage file transfers. In addition, HTML5~\cite{html5} provides a lot of features that enable creating a very powerful, rich and interactive user interfaces. A typical architecture consists of a web client application, a web service to process and provide data, and a storage back-end, which can be a database~\cite{yue2004underlying}.

We designed such a web application that is supposed to visualize climate data interactively in a browser. The client application sends requests to a RESTful web service~\cite{Fielding:2002:PDM:514183.514185}. The web service retrieves data from a NoSQL database, processes them, and sends back the results to the client. For our use case we imported meta data of the Michelson Interferometer for Passive Atmospheric Sounding (MIPAS)~\cite{mipas}, a Fourier transform spectrometer for the detection of limb emission spectra in the middle and upper atmosphere, to the schemaless NoSQL database MongoDB~\cite{Ameri14,mongodb}. As the database contains millions of documents, and the client application expects fast response times to allow for interactive work, there is a need for a horizontally scalable multiprocessing platform, that splits requests and handles them in parallel on a distributed system.

This paper will discuss the design and implementation of this platform to fulfill the before-mentioned requirements. This platform should also help to achieve the main goals of distributed systems, which are, among others, scalability, openness, transparency, manageability and availability~\cite{Tanenbaum:2006:DSP:1202502}. We call the platform Node Scala. It is based on Node.js~\cite{Node01}, a lightweight platform to build scalable network applications, that uses an event-driven, non-blocking I/O model. The 0.10 branch (specifically, version 0.10.33) of Node.js was used instead of the more recent 0.12 because as of now the latter has still largely not been included in standard package repositories of most major Linux distributions, causing possible security concerns for servers running it.

The remainder of this paper is structured as follows. The next section introduces Node.js and explains related work on Node Cluster. In Section~\ref{lab:design} the conceptual details of the design of our platform and its components are described followed by a section on how to extend it with more use cases. In Section~\ref{lab:parallel} the support of different approaches for parallel processing are discussed. The results of an evaluation of our platform are given in Section~\ref{lab:eval} above the conclusion.

\section{Related Work}

\IEEEpubidadjcol

Server applications that implement an event-driven non-blocking model
are more efficient in memory and CPU usage compared to those that
implement a multithreading or multiprocessing model, especially for a
large number of concurrent requests~\cite{C10K}~\cite{Syed:2014:BN:2723784}.

Currently, Node.js is a new technology that implements the
event-driven model to create network
applications~\cite{Cantelon:2013:NA:2601501}. A number of large
Internet services, for instance PayPal~\cite{PayPal} and
LinkedIn~\cite{2014:NLP:2556647.2556656}, have been migrated to
Node.js, resulting in considerable improvements in both performance
and ease of development. For example, in case of PayPal migration from
Java to Node.js improved the system response time by up to 35~\% for
twice the number of concurrent requests per second --- with
development time twice as fast and the size of the code reduced by
33~\%~\cite{PayPal}.

Node.js applications are written in JavaScript and run in a 
non-blocking single thread~\cite{books/daglib/0029097}. In the background
 there is a pool of additional non-blocking threads~\cite{Redkar14}. The 
 main thread and the background threads communicate via queues that 
 represents the assigned tasks. When an I/O task is required, it will be 
 assigned to the background workers. The worker announces the main thread 
 that the task is completed by calling a callback 
 function~\cite{Pasquali:2013:MN:2601491}.

 Each Node.js instance is represented a single
 process~\cite{books/daglib/0029097}. In order to make it possible to
 exploit additional hardware resources of a multi-core host, Node.js
 provides a clustering technology called Node Cluster --- a process
 that spawns multiple child processes on the same machine, each of
 them running a Node.js server instance. These Node.js instances can
 share the same socket~\cite{Node01}. Load distribution between Node
 instances is handled by the underlying operating system, or in
 Node.js version 0.12.0 in a round-robin fashion by the parent
 process. This architecture simplifies managing child processes as
 well as makes it possible to have them automatically restarted in the
 event of a crash~\cite{Node01}. Node Cluster support has been
 incorporated into several advanced Node.js process managers, for
 instance PM2~\cite{PM2} or StrongLoop Process
 Manager~\cite{SLPM}.

 Unfortunately, Node Cluster has several limitations. To begin with,
 the ``defer to the operating system'' scheduling policy used by
 Node.js versions prior to 0.12.0 is known not to be very efficient in
 distributing the load between spawned Node.js
 instances~\cite{StLo01}. Secondly, it can only start Node.js
 instances on a single machine~\footnote{Recent versions of StrongLoop
   Process Manager \emph{are} capable of launching multiple Node
   Cluster instances on different hosts~\cite{SLPMmultihost},
   however this approach requires either an external load balancer
   such as Nginx or developing a custom multi-host
   scheduler. Moreover, instances running on different machines are
   completely unaware of each other.}. Finally, every request sent to
 a Node Cluster application will be processed using one and only one
 thread, regardless of the complexity of the task; there is no
 mechanism in Node Cluster to execute complex tasks in parallel.

In light of the above, we decided to design and implement an
alternative Node.js multiprocessing platform. This platform, which we
have called Node Scala, can have its components distributed across
multiple machines. It also provides a simple mechanism for executing
complex tasks in parallel.

\section{Design}\label{lab:design}

Node Scala should be highly scalable in par with the complexity of
executed tasks and the volume of data. In other words, the scalability
of Node Scala could be measured in terms of complexity of tasks
performed on the used resources. Furthermore, Node Scala should scale
with the number of users of the given resources.

To simplify later improvements of Node Scala, this platform should
fulfill the openness principle~\cite{Tanenbaum:2006:DSP:1202502}
\textit{i.e.} consist of small components communicating with each
other using well-documented protocols and data formats. Such modules
can then be replaced or improved easily without the need to change
other components.

Finally, the system should adhere to the transparency
principle~\cite{Tanenbaum:2006:DSP:1202502}. Namely, any code
incorporated into Node Scala for the purpose of execution of a
specific task should be independent of the number and distribution of
processing components, as well as --- if feasible --- of the
underlying operating system. This of course implies the platform
itself should be system-independent too.

\begin{figure*}
  \centering
  \includegraphics[width=1\textwidth]{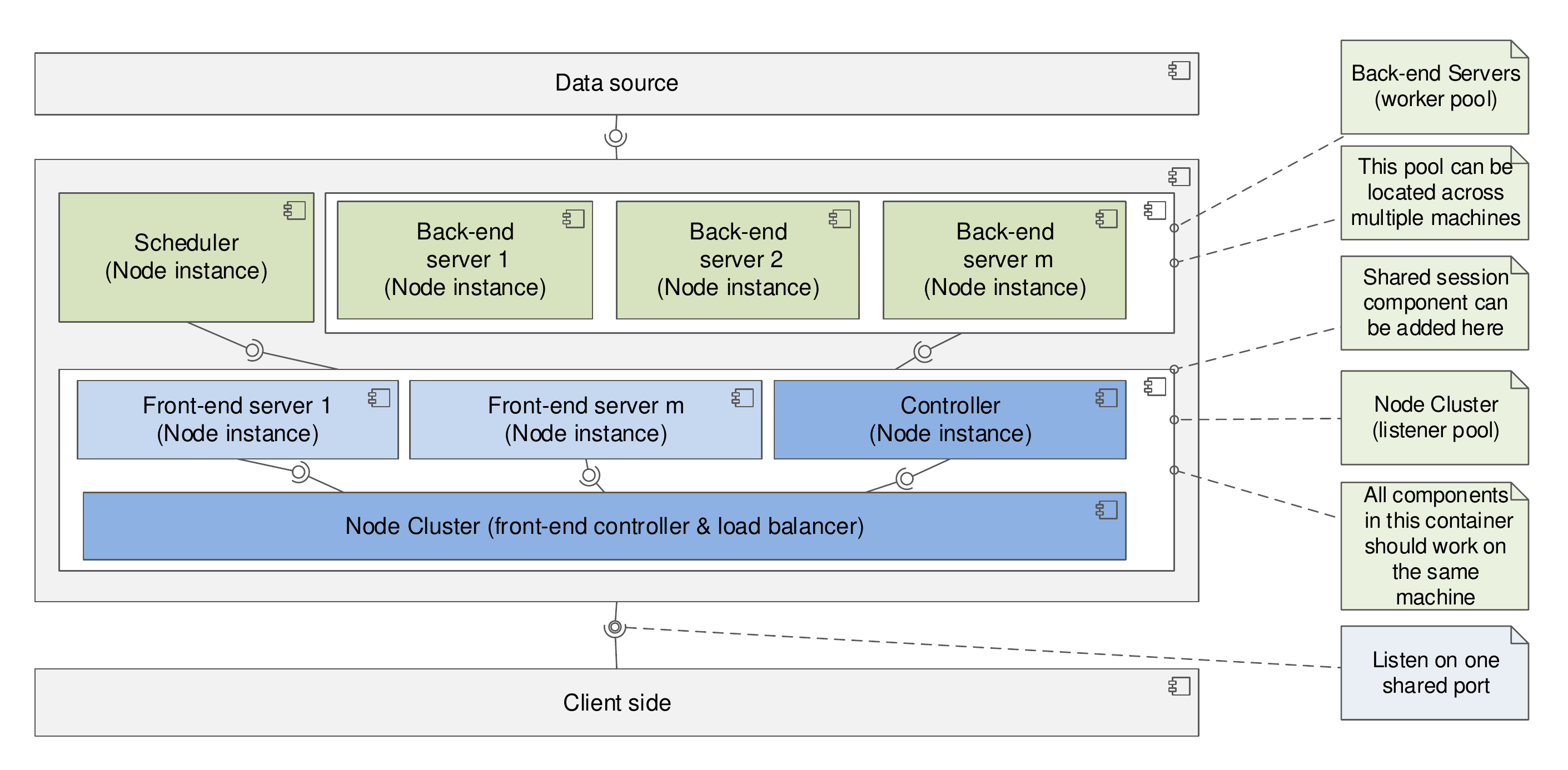}
  \caption{The component diagram of the architecture of Node
          Scala. Each component represents a Node.js instance. The
          components communicate over TCP.}
  \label{fig:Archi_figure}
\end{figure*}

\subsection{The Conceptual Model}

The system consists of front-end servers, back-end servers, the
scheduler and the controller. Front-end servers are responsible for
receiving external requests and dividing them into sub-tasks. Back-end
servers are background workers that execute sub-tasks and return
results as streams. The scheduler is responsible for the distribution
of load between the back-end servers. Finally, the controller handles
start-up and shutdown of other components of the system, even when
they run on multiple machines, as well as monitors and restarts them
as needed to increase overall robustness.

When a request arrives, one of the front-end server receives it and
checks its complexity. If its worth being divided, the front-end
server asks the scheduler for a list of back-end servers and sends one
sub-task to each server. The back-end servers receive the sub-tasks,
execute them and return the result as streams. The front-end server
receives these streams, combines them into one and forwards it to the
caller.

\subsection{The Layers of Background Workers}

Back-end servers of Node Scala are the actual workers, used to execute
expensive tasks. The architecture of these workers can be considered
as consisting of three layers: sub-tasks, back-end servers and the
physical resources. Two mappers are required to handle communication
between these layers. The scheduler is a mapper between sub-tasks and
back-end servers, distributing the former between the latter. The
controller on the other hand acts as a mapper between back-end servers
and physical resources, starting and stopping back-end servers on
interconnected machines -- according to the given configuration file
at start-up time as well as in response to commands issued by the
administrator at run time.

This architecture improves the openness of the system because it makes
it easy to modify or replace individual layer or mapper components.
For example, while the scheduler in this version of Node Scala uses a
simple round-robin algorithm, a drop-in replacement scheduler can be
implemented in the future which provides multiple scheduling
algorithms and the possibility for the administrator to select them as
needed. Another example would be an improved controller, capable of
intelligent allocation and releasing of resources according to the
needs of the system.

\subsection{Management of front-end servers}
Node Cluster serves as load balancer between front-end servers in this version of Node Scala. Additionally, Node Cluster is responsible for starting and stoping the controller and the front-end servers according to the settings 
given in the configuration file. Furthermore, Node Cluster restarts 
front-end servers and the controller in the event of a crash.

Utilizing Node Cluster in this context simplifies sharing in-memory objects 
between front-end servers such as session information by adopting third 
party components designed for Node Cluster, e.g. TLS-session store from StrongLoop team~\cite{StLo02}. However, this way front-end servers can only 
be started only on the same machine of Node Cluster. It is worth mentioning 
that the ideal number of Node instances on a host machine is the number of 
the CPUs in this machine~\cite{Syed:2014:BN:2723784}.

\subsection{System Components}

The component diagram in Figure~\ref{fig:Archi_figure} illustrates the
architecture of Node Scala, with each component representing an
instance of Node.js. A complete Node Scala system consists of one
scheduler, one controller, and at least one front-end server and one
back-end server.

The Node Cluster component visible in the diagram serves as the
first-stage launcher. When the system starts, Node Cluster starts the
Node Scala controller along with the specified number of front-end
servers.  The controller in turn starts the scheduler and the
configured back-end servers.

\subsection{Connecting the Components}

For the sake of transparency of the system, all the components of Node
Scala communicate over TCP network connections. This way connections
between components are independent of their respective locations or
the underlying operating systems.

Data is transferred between components in the JavaScript Object 
Notation (JSON) format~\cite{JSON}, a \textit{de facto} standard of 
data exchange in web applications. The TCP interface in Node.js supports natively
only the transfer of buffers of data and strings~\cite{Node02}. Therefore, 
Node Scala provides a thin layer adding the functionality to transfer objects over TCP.

Back-end servers read the data from the data source. The data is read
as stream. At each request, the data source pushes a data object to
the back-end server. The back-end server processes this object and
forwards it directly to the front-end server using streams. That
means, there is a stream chain that extends from the database to the
client side. Node.js streams are efficient from the point of view of
memory and CPU usage~\cite{Syed:2014:BN:2723784}, avoid caching large
amounts of data in memory and enable transferring and processing data
in parallel. Moreover, the developer can easily extend this chain of
streams to apply complex algorithms or some business logic to the
piped data.

At the end, the front-end server receives multiple streams from
back-end servers. It combines them into one stream and forwards it to
the caller. Corruption of the data as it is received and merged should
of course be avoided. The front-end server can compress the data on
the fly before sending it to the client side, using GZIP compression
algorithm~\cite{GZIP}.

\section{Use cases}

To ensure the reusability of Node Scala, we separated the logic of the
platform itself from the tasks that will be executed in parallel on
this platform, which we refer to as 'use cases'. Use cases can be
defined in the development phase and also while the system is running
without any downtime.

Each use case consists of four attributes: the URL that will be used
from the client side to call this use case, the name of the use case,
the functions that will be executed on the front-end servers and the
functions to be executed on back-end servers. The front-end server
sends the name of the use case and the needed parameters to execute
the specified function on a back-end server. The function on the
back-end server, which is called use-case handler, expects the name of
the use case and its parameters. It returns the results as a
stream. When a back-end server receives the use-case name it forwards
it to the appropriate function, which handles the request and returns
the result to the front-end server.

\section{Parallel-processing Paradigms}\label{lab:parallel}

Node Scala provides a simple interface called ParallelCommander to
simplify dividing and executing commands in parallel. It provides two
paradigms of parallel executing: concurrent parallel and iterative.

\begin{figure*}
  \centering
  \includegraphics[width=1\textwidth]{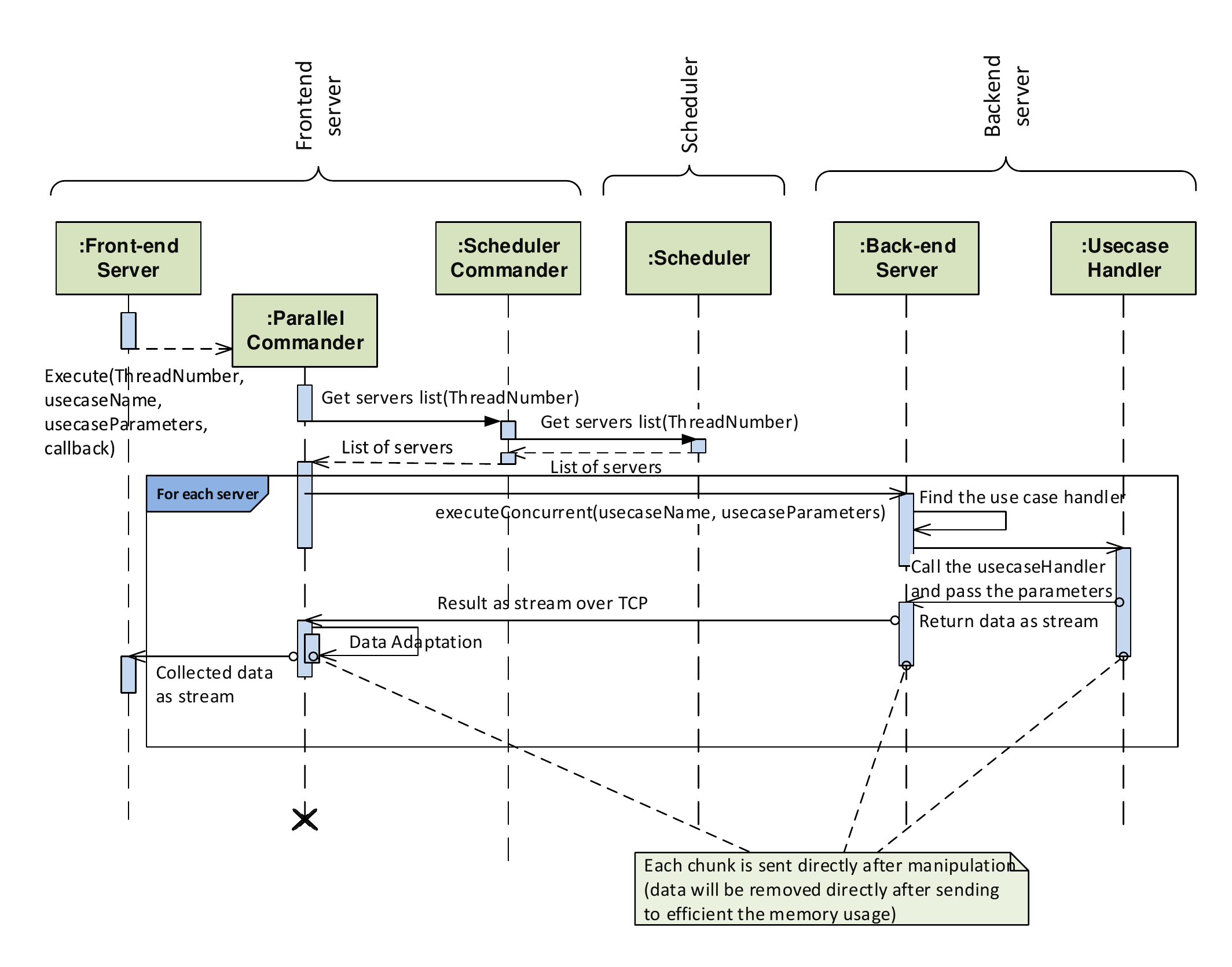}
  \caption{The sequence Diagram of ParallelCommander.executeConcurrent(): execute tasks in parallel concurrently and collect the results in one stream.}
  \label{fig:execute_concurrent}
\end{figure*}

The ParallelCommander interface provides a function called
\texttt{executeConcurrent()} that implements concurrent parallel execution. It
expects a list of sub-tasks and the execution options. The returned
value is a stream that represents the whole result of executing the
sub-tasks. When this function is called, all sub-tasks are executed on
the back-end servers at the same time (see Figure~\ref{fig:execute_concurrent}).

For example, if there are ten sub-tasks to be executed and the system
has five back-end servers, in this model each back-end server will
receive two sub-tasks at the beginning of the execution.

\begin{figure}
  \centering
    \includegraphics[width=80mm]{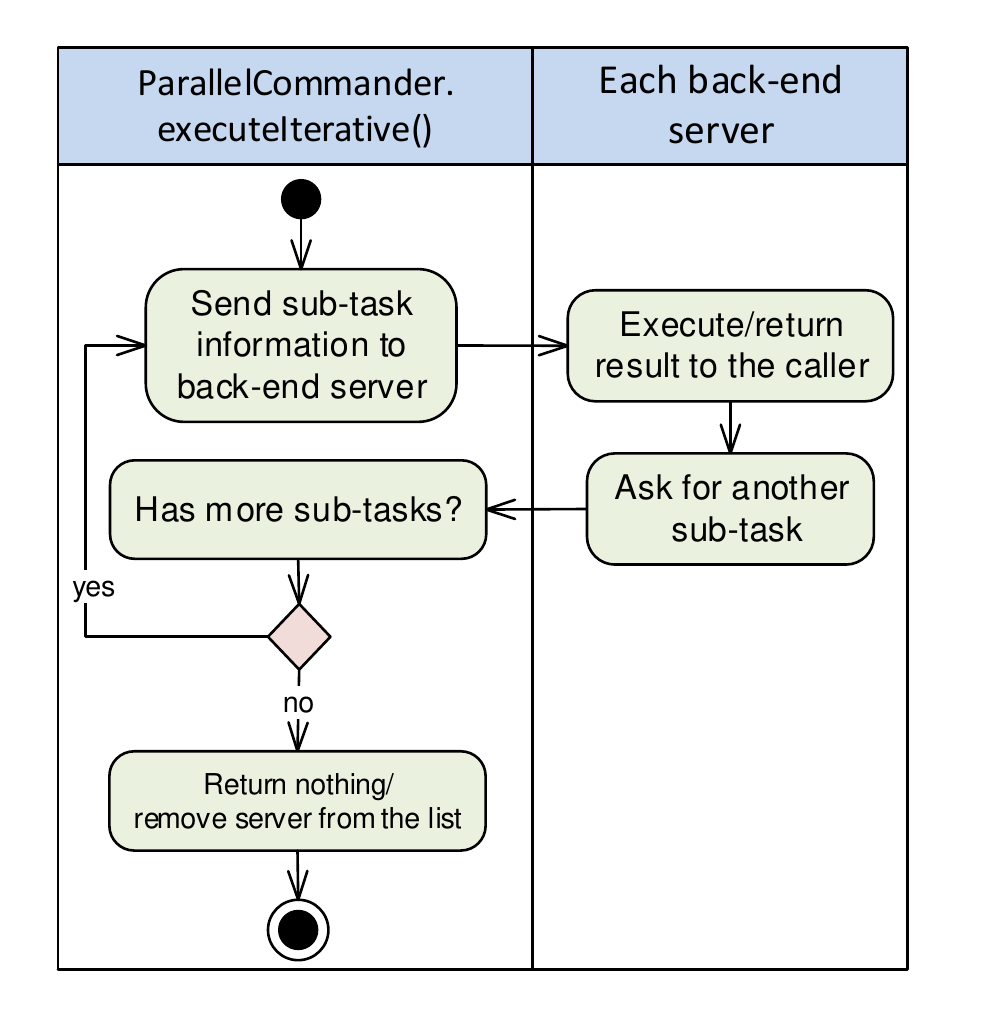}
  \caption{Activity diagram of \texttt{executeIterative()}: execute iterative model to execute tasks in parallel.}
  \label{fig:execute_iterative}
\end{figure}

However, the function \texttt{executeIterative()} of ParallelCommander
allows parallel execution of sub-tasks according to a different
paradigm.  The function expects the number of required back-end
servers and assigns sub-tasks to servers one at a time. When a
back-end server finishes the execution of a sub-task, it asks
ParallelCommander for another. If ParallelCommander has another
sub-task queued, it sends it to the back-end server; otherwise it
sends nothing, indicating to the back-end server that there are no
more sub-tasks to execute (see activity diagram in Figure~\ref{fig:execute_iterative}).

\texttt{executeIterative()} is beneficial for many cases when the
number of expected sub-tasks is very large compared to the number of
available back-end server. For example, imagine that there are
\emph{x} back-end servers and \emph{100x} sub-tasks.  Using
\texttt{executeConcurrent()}, each server receives 100 sub-task requests at the
same time to work on them in parallel. If the system serves more than
one request like this simultaneously, back-end servers might be
overloaded. Instead, \texttt{executeIterative()} can be used with, for
example, \emph{10x} required threads, which means that each server
receives 10 sub-task requests at the same time. When a sub-task is
processed, the back-end server asks for the next sub-task and so
on. Furthermore, this method is very helpful for use cases in which
sub-tasks depend on each other.

\section{Evaluation}\label{lab:eval}

In this section we present the results of evaluating Node Scala as a
horizontally scalable platform.  Furthermore, the impact of tuning
Node Scala settings on performance is discussed as well. The use case
used during evaluation is a real-world application that provides a
RESTful web service for executing complex algorithms on climate data,
an example of Big Data.

All performance measurements are taken from the client side. The
response time represents the time period between sending the URL
request and receiving the whole result.

\subsection{Data Description}

The data is saved in MongoDB and represents metadata of measurements of climate
data such as concentration of trace gases, pressure and temperature in
the atmosphere. These measurements were taken by MIPAS, an interferometer mounted on the European environmental satellite ENVISAT. MIPAS/ENVISAT operated from 2002 to 2012. The corresponding metadata consists of about 5.3 million documents~\cite{Ameri14}. Each document corresponds to the measurements at a specific geolocation and
time. The considered data are modelled as a three-dimensional array for
the data attribute, measurement number and the used channel. Each
channel represents an instrument. In this data set, three channels are
used. This array contains the cloud-index values, a measure for the formation of clouds in the sky.

\subsection{Data-processing Algorithm}

In this use case each document is processed by a CPU-intensive
algorithm and filtered according to a given channel and a threshold
cloud-index value to provide cloud altitude at given point and
time. The format of the output objects can be selected by the received
query and grouped by days. Besides the channel and threshold value,
the required date range has to be specified.

The following example gives an idea about the processing size: when
the client asks for the data of the month 07/2003, the number of
selected documents is about 61,000 and the size of returned data is
about 10~MB (changed depending on the used format and given
parameters). Thus, the data is compressed on the fly before
transferring to the caller using the GZIP~\cite{GZIP} algorithm, which
is supported by most web browsers.

In order to guarantee efficient memory usage, receiving and processing
data is implemented as streams which extend the Node Scala stream
chain. Compressing the data before sending it to the client is also
achieved using streams.

\subsection{Test Environment}

The client and the servers run on two separate machines. Each machine
has two Intel(R) Xeon(R) E5-2640 CPUs, running at a frequency of
2.5~GHz. Each CPU has 6 cores and feature Hyper-Threading, meaning
there are 24 VCPUs available per machine. The operating system on the
servers is Linux, specifically CentOS~6 (64-bit version). Each server
has 128~GB of RAM. The two servers are located on the same physical location and
connected directly without any firewalls or proxies between them.

In the following experiments, Node.js version 0.10.33 and MongoDB version 2.6.7 were used.

Please note that that in all following tests all Node Scala components
operated on the same machine in order to facilitate comparison to Node
Cluster. On the other hand, the MongoDB server runs on the
client-machine, \textit{i.e.} not on the same host as Node Scala.

\subsection{Performance-measurement Tools}

To emulate the web browser, Apache JMeter~\cite{JMeter} (version
v2.12) is used, which is designed to test web applications and measure
their performance. It registers the elapsed time between sending the
URL and receiving the whole required data. The experiment results can
be saved in multiple formats to be analysed by various tools.

A Node.js application has been implemented in order to measure memory
and CPU consumption. This application uses the 'top' Linux command, in
its batch mode, to collect information about resource usage at a given
interval (two seconds in the following experiments) and saves the
result in CSV (Comma-Separated Values) format.

\subsection{Comparing Scaling Behaviour of Node Scala to Node Cluster}

\begin{figure}
  \centering
    \includegraphics[width=80mm]{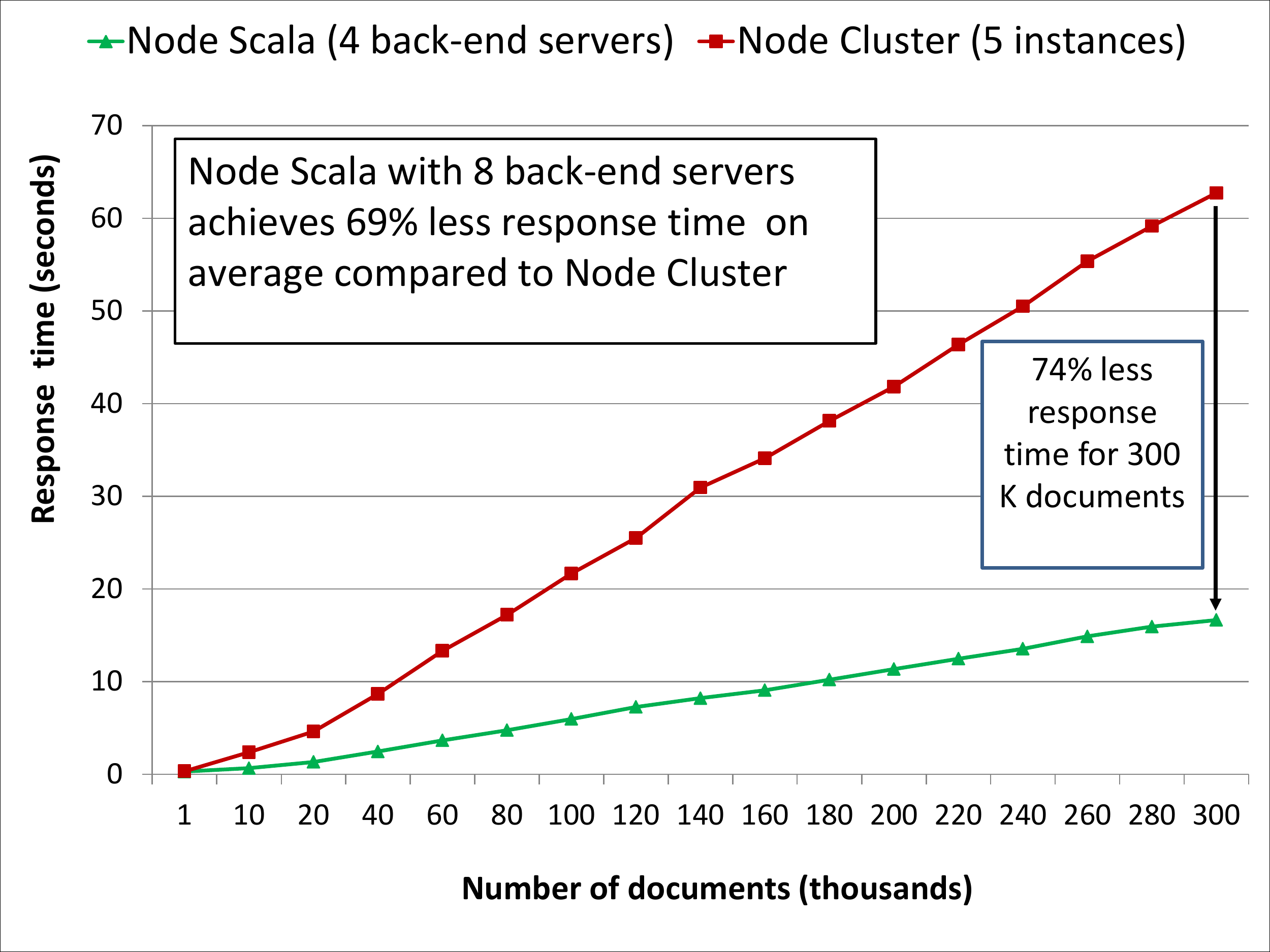}
  \caption{Comparing scalability of Node Scala to Node Cluster.}
  \label{fig:cluster_scala_performance}
\end{figure}

The first experiment, shown in
Figure~\ref{fig:cluster_scala_performance}, compares Node Scala to
Node Cluster (and by extension PM2, StrongLoop Process Manager and
other similar tools which use Node Cluster internally) in terms of
scalability. In this experiment only one request was issued at a
time. Node Cluster has five Node.js instances and Node Scala has one
front-end server and four back-end servers. This experiment is
repeated 60 times. The average of the measured response times is
calculated and considered as the final result. As illustrated in
Figure~\ref{fig:cluster_scala_performance}, the response time of Node
Scala is much smaller than Node Cluster for all values of the number
of documents in the tested range of 1,000 to 300,000. As the number of
requested documents grows the difference in response time of the two
platforms grows; for the highest measured number the response time of
Node Scala is 74~\% of that of Node Cluster. In other words, the more
complicated the request the greater the benefit of using Node Scala.

\subsection{Increasing the Number of Back-end Servers}

\begin{figure}
  \centering
    \includegraphics[width=80mm]{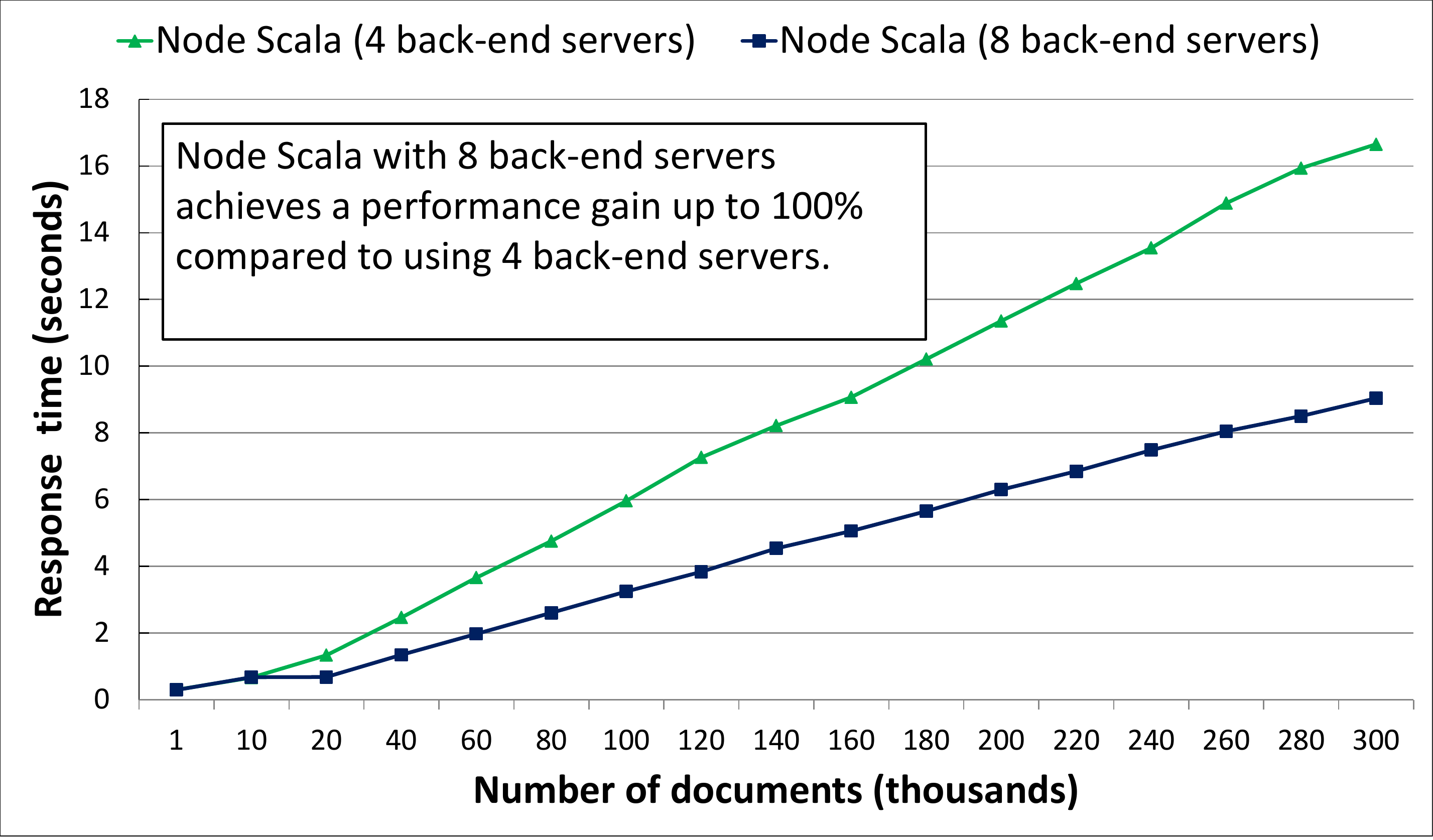}
  \caption{The influence of increasing resources (back-end servers) on the performance of Node Scala approach.}
  \label{fig:node_scala_resource_expensive}
\end{figure}

The second experiment (see
Figure~\ref{fig:node_scala_resource_expensive}) illustrates horizontal
scalability of Node Scala by increasing the number of processing
resources from four to eight back-end servers. When the number of
documents is small, the performance of both configurations is almost
the same. However, when the number of requested documents
(\textit{i.e.} the complexity of the requested task) grows and the
response time increases accordingly, doubling the number of back-end
servers can result in performance gain of up to 100~\%. As expected,
for a low number of documents to be processed the overhead of
splitting the task causes a reduced speed-up.

\subsection{More Concurrent Requests}

\begin{figure}
  \centering
    \includegraphics[width=80mm]{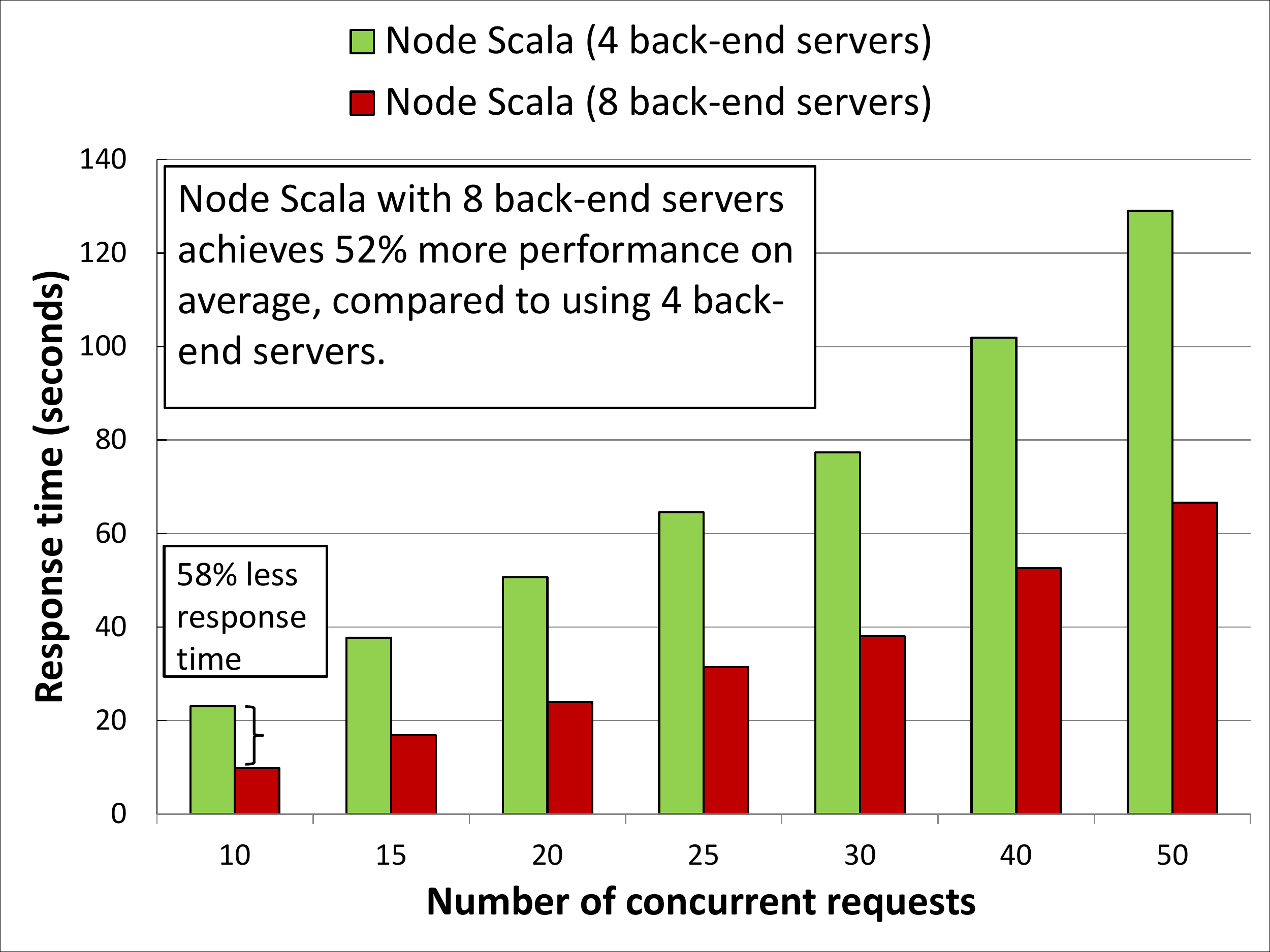}
  \caption{Response time in Node Scala approach when the number of concurrent requests and number of users increase.}
  \label{fig:Scala48_con}
\end{figure}

All the previous measurements were taken for one request at a
time. Obviously, when the number of concurrent requests increases the
system's response time per request will be
longer. Figure~\ref{fig:Scala48_con} shows the influence of the number
of concurrent requests on the system performance.

The request of processing data in one month is the most frequent query
in this use case. Therefore, each request in this experiment asks for
60,000 documents, which represents the average number of documents in
one month.

When the number of requests increases, the response time of Node Scala
increases significantly. However, if the number of back-end servers
increases, Node Scala will have shorter response time as a consequence
of distributing tasks on the additional
servers. Figure~\ref{fig:Scala48_con} shows how Node Scala with 8
back-end servers achieves up to 58~\% performance gain compared to 4
back-end servers.

The experiment shown in Figure~\ref{fig:Scala48_con} demonstrates how
our system scales with the number of allocated processing
resources. The results show the ability of the Node Scala approach to
scale horizontally even with increasing the number of concurrent
requests (number of users).

Only one front-end server is used for the two systems. They differ
only in the number of back-end servers.

This experiment was repeated 20 times. The initialization and
finalization experiments were excluded then the average of these
experiments are collected and registered as the final result.

\subsection{Resource usage profile}

Table~\ref{tab:cluster_resources} explains the average of CPU and
memory usage of Node Cluster while the system is responding to the
requests in a stress test. This table demonstrates the hitherto
mentioned inefficiency in distributing the load between Node.js
instances in Node Cluster prior to version 0.12.0~\cite{StLo01}: in
one experiment, one Node.js instance is almost overloaded, while some
other instances used less than 50~\% from the allocated CPU. The new
scheduling policy introduced and made default in Node.js branch 0.12
is expected to make resource usage of Node Cluster more efficient,
however given the prevalence of older versions of Node.js on
production-grade Web servers it will take some time before the
shortcomings of the old policy can be considered irrelevant.

\begin{table}
  \caption{Memory and CPU usage of Node Cluster with five instances during stress test (two experiments).}
  \label{tab:cluster_resources}
  \centering
    \begin{tabular}{|c|c|c||c|c|}
      \hline
      \multirow{2}{*}{\textbf{Instance No.}} & \multicolumn{2}{|c|}{\textbf{CPU\%} }& \multicolumn{2}{|c|}{\textbf{Memory\%} } \\
      \cline{2-5}
      & \textbf{ 1. Exp. } & \textbf{ 2. Exp. } & \textbf{ 1. Exp.} & \textbf{ 2. Exp.} \\
      \hline
      1 & 84.0   & 26.0 & 9.9E-02 & 9.6E-02 \\
      2 & 17.9 & 99.1 & 8.0E-02 & 1.0E-01 \\
      3 & 71.8 & 61.7 & 9.9E-02 & 9.3E-02 \\
      4 & 49.5 & 62.2 & 9.4E-02 & 9.2E-02 \\
      5 & 46.5 & 35.6 & 9.8E-02 & 9.9E-02 \\
      \hline
    \end{tabular}
\end{table}

Table~\ref{tab:scala_resources} summarizes the resource usage of Node
Scala and shows its efficiency in distributing the load between
back-end servers. In all experiments the gathered resource data is
almost the same. Furthermore, this table shows how the front-end
server efficiently works as a light proxy. Front-end server is a thin
layer between the client side and the back-end servers that divides
and distributes tasks and then forwards the result as streams back to
the client without using a lot of processing resources, which enable
it to receive new requests and handle them quickly. Front-end servers
can also serve simple tasks. However, front-end servers should not
execute an CPU-intensive tasks to avoid degrading the system
availability; CPU-intensive tasks should always be executed on the
back-end servers.

The scheduler and the controller can be considered almost as an idle
process. In order to improve the system availability in future
versions of Node Scala, the scheduler can be attached to the front-end
server. In this case, the front-end server and the scheduler will
share the same CPU code. The more front-end servers the system has,
the more schedulers the system has, thus removing a single point of
failure from the architecture. This feature can easily be implemented
by having schedulers started by respective front-end servers.

Thanks to streams and the event-driven model, the memory usage is
efficient in both the Node Cluster and the Node Scala approaches. In
spite of increasing the number of concurrent requests or the cost of
the tasks, the memory usage remains almost constant.

\begin{figure}
  \centering
    \includegraphics[width=80mm]{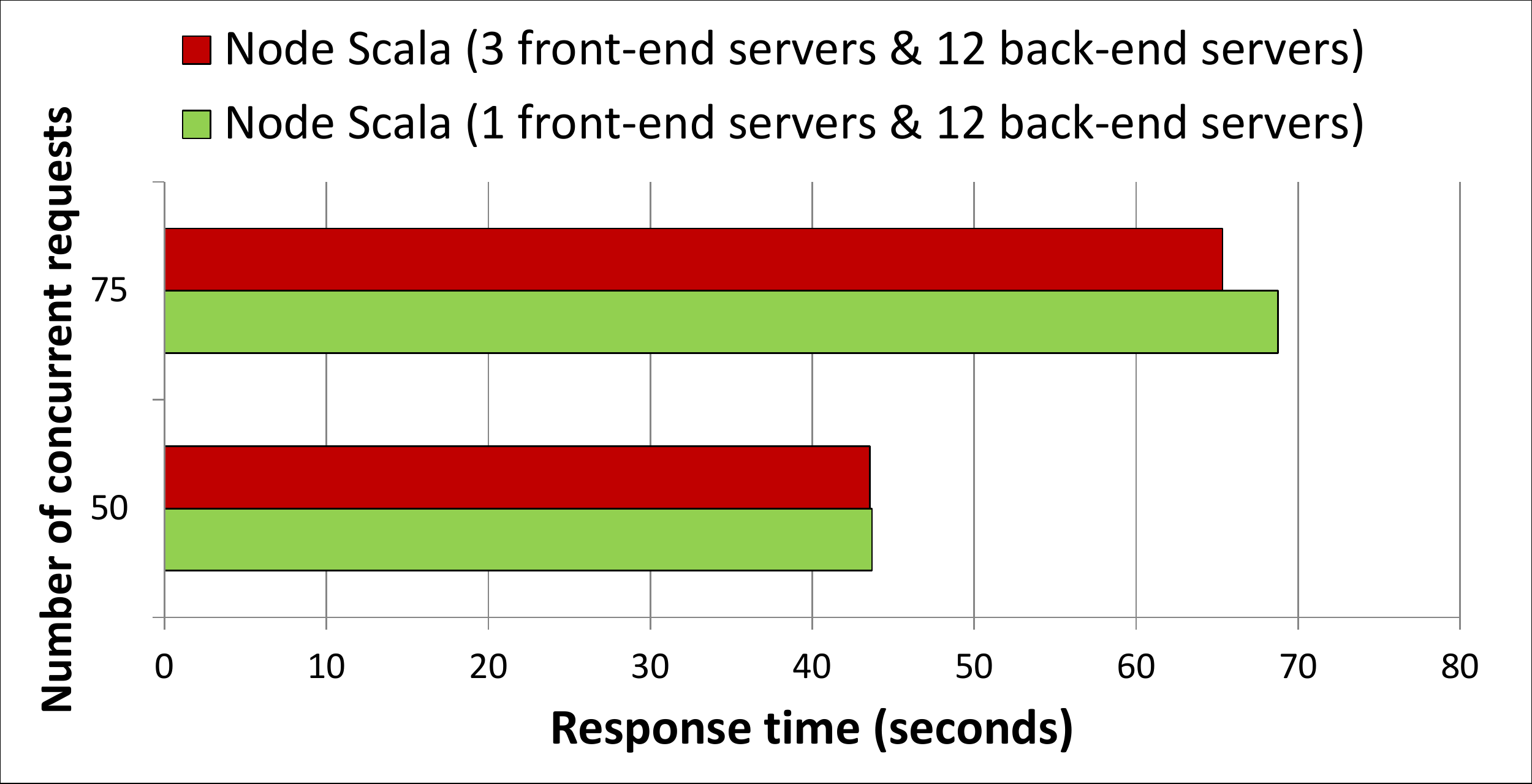}
  \caption{The influence of increasing front-end servers for large number of concurrent requests.}
  \label{fig:increase_fs}
\end{figure}

Figure~\ref{fig:increase_fs} shows the slight effect on performance of
adding another front-end server to a Node Scala system. When the
number of concurrent requests increases, front-end servers help to
reduce the response time a bit.

The test results of the evaluation demonstrated the scaling behavior of Node Scala as function of available backend servers and under heavy load coming from concurrent requests. The desired load balancing is achieved and the platform can be easily extended according to the needs of the use-cases.

\begin{table}
  \centering
  \caption{Resource usage in Node Scala with one front-end server and four back-end servers during stress test.}
  \label{tab:scala_resources}
    \begin{tabular}{|l|c|c|}
      \hline
      \textbf{Component} & \textbf{CPU\%} & \textbf{Memory\%} \\
      \hline
      Front-end server  & 0.9 & 9.9E-02 \\
      \hline
      Back-end server 1 & 96.6 & 9.6E-02 \\
      \hline
      Back-end server 2 & 99.6 & 9.9E-02 \\
      \hline
      Back-end server 3 & 97.1 & 9.8E-02 \\
      \hline
      Back-end server 4 & 96.0 & 9.6E-02 \\
      \hline
      Controller & 0.1 & 0.0 \\
      \hline
      Scheduler & 0.02 & 0.0 \\
      \hline
    \end{tabular}
\end{table}

\section{Conclusion}

A multiprocessing platform for Node.js called Node Scala has been
developed which allows for efficient distribution of application tasks
between both multiple cores on a single host and multiple hosts in a
cluster, as well as to perform complex tasks in parallel. The platform
is highly modular, portable and configurable, facilitating deployment
and further development. Performance evaluation conducted using a
real-world use case of processing and serving large amounts of data
stored in a NoSQL database demonstrated consistently smaller response
times comparing to a solution based on Node Cluster, clear horizontal
scalability with the number of available worker nodes, and efficient
use of underlying computing resources. A Node Scala-based RESTful
service is already in use which processes and serves Envisat MIPAS
data to a visualization and online-analysis application.

We are planning to publicly release the source code of Node Scala,
under an Open Source licence, in the near future.

\section*{Acknowledgment}
This work is funded by the project ``Large-Scale Data Management and Analysis''~\cite{jung2013oodlc} funded by the German Helmholtz Association.


\bibliographystyle{IEEEtran}

\bibliography{references}

\end{document}